\documentclass[a4paper,showpacs,twocolumn]{revtex4}
\usepackage{graphicx}
\usepackage{amsmath}
\usepackage{amssymb}
\usepackage{bm}
\usepackage[final]{showkeys} 
\usepackage{hyperref}
\usepackage[T1]{fontenc}
\usepackage[english]{babel}
\usepackage{psfrag}

\def\leqq{\lambda_0^\mathrm{eq}}
\def\lin{\lambda_0^\mathrm{in}}

\def\lam{\lambda_0}
\def\Lam{\Lambda}
\def\ldot{\dot{\lambda_0}}
\def\th{\theta}
\def\thin{\theta_\mathrm{in}}

\def\dens{n} 
\def\cc{ c_2\dens } 

\def\x{ \bm{x} } 
\def\tt{ \nu_0 t }

\def\Tmf{ T_\textrm{mf} }
\def\TZ{ T_\textrm{Zeeman} }

\def\al{\alpha}
\def\mag{ {\cal M} }
\newcommand{\ovl}[1]{ \overline{ #1 } }
\def\Ehfs{ E_\mathrm{hfs} }

\def\Rtf{R_\mathrm{TF}}
\def\nmax{\dens_\mathrm{max}}

\def\sn{\textrm{sn}}
\def\cn{\textrm{cn}}
\def\dn{\textrm{dn}}

\begin{document}

\title{Spin dynamics and structure formation in 
  a spin-1 condensate in a magnetic field} 

\author{Jordi Mur-Petit}
\email{ j.mur@ucl.ac.uk }
\affiliation{ Dept.\ Physics and Astronomy, UCL, Gower Street, WC1E
  6BT, London, UK, EU }

\date{ \today }


\begin{abstract}

We study the dynamics of a trapped spin-1 condensate in a
magnetic field. 
First, we analyze the homogeneous system, for which the dynamics can
be understood in terms of orbits in phase space.
We analytically solve for the dynamical evolution of the populations
of the various Zeeman components of the homogeneous system.
This result is then applied via a local density approximation to
trapped quasi-1D condensates.
Our analysis of the trapped system in a magnetic field shows that both
the mean-field and Zeeman regimes are simultaneously realized, and we
argue that the border between these two regions is where spin domains
and phase defects are generated. We propose a method to
experimentally tune the position of this border.

\end{abstract}

\maketitle

\section{Introduction}

Bose-Einstein condensates (BECs) with a spin degree of freedom are
an interesting field of research in many-body physics as they realize
both superfluidity and magnetism in a well-controlled environment.
First realized experimentally with $^{23}$Na ten years
ago~\cite{StamperKurn1998,Stenger1998}, their study has matured
remarkably over the last few years, with several groups studying their
dynamics~\cite{Chang2004,Schmaljohann2004,Kronj2005,Higbie2005} and
thermodynamics~\cite{Erhard2004,Schmaljohann2004apb}.
Of particular interest is the study of the process by which spin
domains are formed during time evolution, a phenomenon observed
experimentally~\cite{Chang2005,Higbie2005,Sadler2006} and in numerical
simulations based on a mean-field
approach~\cite{Mur2006,Saito2005,Zhang2005prl}.

The complicated dynamics of these non-linear systems, especially when
they are subjected to time-varying external fields, makes the physical
understanding of the structure formation process somehow elusive.
To address this point, we present here a simple model based on an
analytic solution for the homogeneous system for arbitrary
magnetic fields $B$ and magnetizations $\mag$.
This solution is then applied to the study of realistic, trapped
spin-1 condensates by means of the local density approximation
(LDA). This approximation has already been applied successfully in a
number of studies on scalar BECs, as well as cold Fermi gases.
From the analysis of our results we are able to provide an intuitive
picture of the process leading to the structure formation.
Further, we argue that it should be possible to experimentally
``tune'' the spatial region where this process starts within the
condensate.

The paper is organized as follows.
In Sect.~\ref{ssec:static} we present the phase space of a homogeneous
system under a magnetic field $B$ and for arbitrary $\mag$, and
introduce the phase-space orbits that describe the dynamics of a
conservative system.
In Sect.~\ref{ssec:dynam}  we solve analytically the dynamical
evolution of the homogeneous system.
Then, in Sect.~\ref{sec:dynam} we describe our local-density
approximation for a trapped system  and present numerical results for
its dynamics (Sect.~\ref{ssec:lda}), which we compare with simulations
based on a mean-field treatment (Sect.~\ref{ssec:gpe}). 
In Sect.~\ref{sec:dephasing} we discuss the progressive dephasing of
different spatial points of the condensate in a homogeneous magnetic
field, and relate this to the process of structure formation, with an
indication of a possible experimental test.
Finally, we conclude in Sect.~\ref{sec:conclusions}. 

\section{Analytical results for the homogeneous system}
\label{sec:homog}

\subsection{Energetics of the homogeneous system}
\label{ssec:static}

A homogeneous condensate of atoms with total spin $F$ can be described 
by a vector order parameter $\vec{\psi}$ with $2F+1$ components,
\begin{align}
  \vec\psi &= \left(
    \begin{array}{c}
      \psi_{F} \\
      \vdots \\
      \psi_{-F}
    \end{array}
    \right) \:.
    \label{eq:order}
\end{align}
The density of atoms in a given Zeeman component $m=-F,\cdots,F$ is
$\dens_m = |\psi_m|^2$ and 
the total density is given by $\dens=\sum_m|\psi_m|^2$.
Introducing the relative densities for the homogeneous system
$\lambda_m = \dens_m/\dens$, one has
\begin{align}
  \sum_m \lambda_m = 1 \:.
  \label{eq:norm}
\end{align}
Given that $\dens$ is a conserved quantity, Eq.~(\ref{eq:norm}) will
be fulfilled at all times during the dynamical evolution.
Moreover, the magnetization
\begin{align}
  \mag = \sum_m m\lambda_m
  \label{eq:mag}
\end{align}
is also a conserved quantity~\cite{Mur2006}.

\begin{figure}
  \centering
  \includegraphics[width=0.3\textwidth]{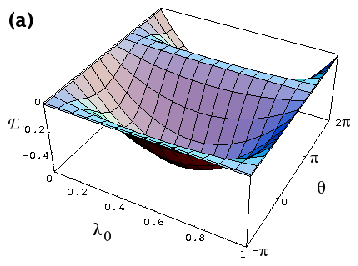}\hfill%
  \includegraphics[width=0.3\textwidth]{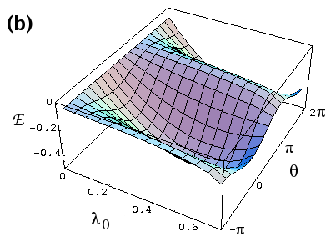}
  \caption{
    Energy (in units of $|c_2|\dens$) of the homogeneous system for
    the cases (a) $\mag=0,~B=0$ and (b) $\mag=0.3,~ B=1$~mG, as given
    by Eq.~(\ref{eq:Ehomog}) for a spin-1 condensate of 
    $^{87}$Rb.}
  \label{fig:phase_space}
\end{figure}

We now focus our analysis to the case of a $F=1$ condensate.
We write 
the various components of the order parameter
as
$\psi_m = \sqrt{\dens\lambda_m}\exp(i\theta_m)$.
This ansatz, together with conditions (\ref{eq:norm}) and
(\ref{eq:mag}), leads to the following expression for the
energy per particle of the homogeneous system in the mean-field
approach~\cite{Moreno2007,Zhang2003}:
\begin{align}
  {\cal E} (\lam,&\mag, \th) = 
  c_2\dens\Big[ \lam(1-\lam) + \frac{\mag^2}{2} + \nonumber \\
    & + \lam\sqrt{(1-\lam)^2 - \mag^2}\cos\th \Big] 
    + \delta(1-\lam) 
  \label{eq:Ehomog} \:.
\end{align}
Here $\th=2\th_0-\th_1-\th_{-1}$, while $c_2$ is given in terms of the
$s$-wave scattering lengths $a_f$ in the channels of total spin
$f=0,2$, by $c_2=4\pi\hbar^2(a_2-a_0)/(3M)$, with $M$ as the atomic
mass.
Finally, $\delta=(E_- + E_+ -2E_0)/2$, where the energies of the
atomic Zeeman states are given by the Breit-Rabi
formula~\cite{BreitRabi}
$ E_{m} = -\Ehfs/8 -\Ehfs\sqrt{1+m\al+\al^2}/2$
($m=-1,0,+1$),
with $E_\textrm{hfs}$ being the atomic hyperfine splitting
and $\al= (g_I\mu_N + g_J\mu_B)B/E_\textrm{hfs}$ is a function of the
external magnetic field $B$.
Here, $g_I, g_J$ are the nuclear and electronic Land\'e factors,
and $\mu_N,\mu_B$ are the nuclear and Bohr magnetons, respectively.
A sketch of the surface ${\cal E}$ is given in
Fig.~\ref{fig:phase_space}.

\begin{figure}
  \centering
  \includegraphics[width=0.45\textwidth,clip=true]{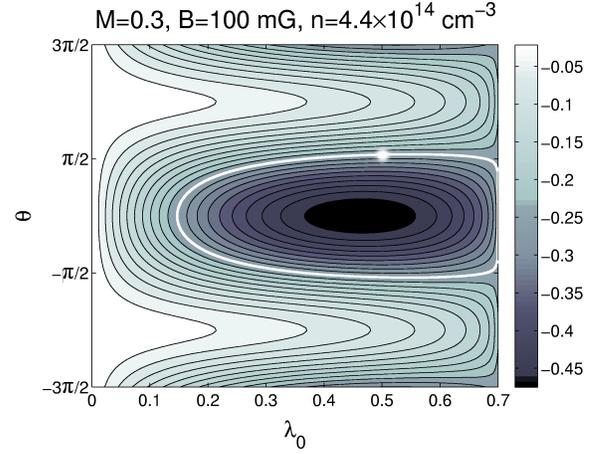}
  \caption{
    Contour plot of the energy surface corresponding to
    $\mag=0.3$ and $B=100$~mG. 
    The white line shows the orbit corresponding to the initial
    conditions $\lin=1/2,~\thin=\pi/2$ (indicated by the white dot).
    The minimum of $\cal E$ is at $\lam\approx0.455,~\th=0$.
    Note the presence of open orbits for energies above that of the
    indicated white line.} 
  \label{fig:orbites}
\end{figure}

As indicated above, $\mag$ is a constant during dynamical
evolution. Similarly, given initial conditions $(\lin, \thin)$,
$E={\cal E}(\lin,\mag,\thin)$ will also be conserved, thus defining an
orbit on the surface ${\cal E}$ in $(\lam,\th)$ space.
A sketch of one such orbit is presented in Fig.~\ref{fig:orbites}.
One should note that, depending on the initial conditions, the orbit 
defined by $E=\textrm{const}$ can be {\em closed} or {\em open}.
In the first case, $\th=\th(t)$ will be a periodic function of time,
while in the latter case, $|\th(t)|$ will grow indefinitely with time. 
In both cases, however, $\lam=\lam(t)$ will be a periodic function of
time. 

\subsection{Dynamics of the homogeneous system}
\label{ssec:dynam}

We are interested in the time evolution of the densities of the
different Zeeman components, $\dens_m=\dens\lambda_m$.
From Eqs.~(\ref{eq:norm}) and~(\ref{eq:mag}) we have that 
\begin{align}
  \lambda_{\pm1}   &= \frac{1 \pm \mag - \lam}{2} \:.
  \label{eq:lambdes}
\end{align}
Therefore, we only need to follow the evolution of $\lam$, which is given by
\begin{align}
  \frac{\partial \lam}{\partial t} \equiv \dot{\lam}
  = \frac{2}{\hbar} \cc \lam\sqrt{(1-\lam)^2 - \mag^2}\sin\th
  \:.
\end{align}
With Eq~(\ref{eq:Ehomog}), we rewrite this as
\begin{align*}
  &(\ldot)^2 
  =
  \frac{4}{\hbar^2}\bigg\{
  (\cc\lam)^2\left[(1-\lam)^2 -\mag^2\right] 
  \nonumber \\
  &-\left[ E -\delta(1-\lam) -\cc\left(\lam(1-\lam)+\frac{\mag^2}{2} \right) 
    \right]^2 \bigg\}  
  \:.
\end{align*}
It can be shown that the term in $\lam^4$ actually drops out and we
are left with a cubic polynomial on $\lam$,
\begin{align}
  (\ldot)^2 
  & 
  \equiv A(\lam-\Lam_1)(\lam-\Lam_2)(\lam-\Lam_3) \:,
  \label{eq:rhodot}
\end{align}
with
\begin{equation}
  A :=-\frac{8\cc\delta}{\hbar^2} \:,
  \label{eq:A}
\end{equation}
and $\Lam_j~(j=1,2,3)$ are the roots of $(\ldot)^2$, 
$\Lam_1 < \Lam_2 < \Lam_3$.
For ground state ($F=1$) alkalies $\delta>0$. 
Therefore, $\Lam_1 \leq \lam(t) \leq \Lam_2$ for $c_2<0$
and $\Lam_2 \leq \lam(t) \leq \Lam_3$ for $c_2>0$~\cite{Zhang2005}.
For concreteness, in the following we will assume $c_2\leq 0$, i.e.,
ferromagnetic interactions.

We will now integrate the time evolution of $\lam$.
To do so, we introduce an auxiliary variable $z$
through $\lam = (\Lam_2 - \Lam_1)z^2 + \Lam_1$. This will satisfy the
differential equation
\begin{align}
  \dot{z} 
  = \frac{\sqrt{A}}{2}\sqrt{\Lam_2-\Lam_1} \sqrt{ (z^2-1)(z^2-k^{-2}) }
  \label{eq:zdot}
\end{align}
where we defined
\begin{equation}
  k^2 := \frac{\Lam_2-\Lam_1}{\Lam_3-\Lam_1} \in [0,1] \:.
  \label{eq:k}
\end{equation}
The first order differential equation~(\ref{eq:zdot}) can be solved
analytically by separating the variables $z$ and $t$, and integrating:
\begin{align*}
  \frac{\sqrt{A}}{2} \int_{t_0}^t dt
  &=
  \frac{1}{\sqrt{\Lam_2-\Lam_1}} \int_{z_0}^{z_t} 
  \frac{dz} { \sqrt{ (1-z^2)(k^{-2}-z^2) } } =
  \\
  &=
  \frac{1}{\sqrt{\Lam_3-\Lam_1}} \int_{z_0}^{z_t} 
  \frac{dz} { \sqrt{ (1-z^2)(1-k^2z^2) } } \:.
\end{align*}
The solution to the last integral can be expressed
in terms of the elliptic integral of the first kind~%
\footnote{
  Note that some references define $F(\phi,k)$ as
  $ F(\phi, k) =\int_0^{\sin\phi} dz / \sqrt{ (1-z^2)(1-kz^2) }$.
},%
\begin{equation*}
  F(\phi, k) = \int_0^{\sin\phi} 
   \frac{dz}{ \sqrt{ (1-z^2)(1-k^2z^2) } } \:.
\end{equation*}
Taking as initial condition $z(t=t_0)=z_0$
and using the fact that $F(-u, k)=-F(u, k)$,
we can express $z_t$ in a compact form by means of the
Jacobi elliptic functions~\cite{Abramowitz}, which are defined
as the inverses of the elliptic integrals,
\begin{align}
  z_t \equiv z(t) = \sn\Big(\gamma_0 + 
  \frac{\sqrt{A(\Lam_3-\Lam_1)}}{2}(t-t_0) ~\Big|~ k \Big)
\end{align}
with $z_0 = \sn(\gamma_0|k)$, i.e., $\gamma_0 := F(\arcsin(z_0), k)$.
Finally, we undo the change in variables to write down the time
evolution of the population of the $|m=0\rangle$ state,
\begin{align}
  \lam&(t) 
  = 
  \Lam_1 + \nonumber \\
  &(\Lam_2-\Lam_1)\,
    \sn^2\!\left(\gamma_0 + 
           \frac{\sqrt{A(\Lam_3-\Lam_1)}}{2}(t-t_0)~%
           \Big| k \right) .
  \label{eq:rhot}
\end{align}
In accordance with the identity~\cite{Abramowitz}
\begin{equation*}
  \sn^2(\al | k) = \frac{1-\cn(2\al | k)}{1+\dn(2\al | k)} \:,
\end{equation*}
and given that both $\cn(2\al|k)$ and $\dn(2\al|k)$ are periodic functions
in $\al$ with period $2K(k)$, $\lam(t)$ will be a periodic function of time
with period 
\begin{equation}
  \label{eq:period}
  T = \frac{2\hbar}{\sqrt{-2\cc\delta(\Lam_3-\Lam_1)}}
    \, K\left( \sqrt{ \frac{\Lam_2-\Lam_1}{\Lam_3-\Lam_1} } \right)  \:.
\end{equation}
Here, $K(k)=F(\pi/2, k)$ stands for the complete elliptic integral of 
the first kind.
We note that result (\ref{eq:period}) agrees with that in
Ref.~\cite{Zhang2005}, where $T$ was calculated directly by performing 
the integral $T=\oint d\lam/\ldot$ over a period of evolution.
Further, let us point out that the average value 
$ \lam^\textrm{av} = (1/T)\int_{t_0}^{t_0+T} \lam(t)\,dt$
does not necessarily coincide with the position of the minimum of
$\cal E$, i.e., $\lam^\textrm{av}$ may differ from the 
equilibrium value $\leqq$ 
(as given, e.g., in Ref.~\cite{Moreno2007} for the case $B=0$).
This is illustrated in Figs.~\ref{fig:rhot}(b)
and~\ref{fig:orbites2}(b).

\subsection{Evolution in the absence of a magnetic field}
We observe that the representation of $(\ldot)^2$ as a cubic polynomial
on $\lam$, Eq.~(\ref{eq:rhodot}), cannot be performed when $A=0$, i.e.,
when $B=0$. In this case, the analytic expression~(\ref{eq:rhot})
is meaningless, as it would apparently result in no time evolution at
all. Actually, in this situation, $(\ldot)^2$ can be written as a
quadratic polynomial on $\lam$:
\begin{align}
  (\ldot)^2   
  &=
  -\frac{4}{\hbar^2}\left\{ 2\cc E\lam^2
    -2\cc E'\lam +(E')^2  \right\}
    \nonumber \\
  &\equiv \ovl{A}(\lam-\ovl{\Lam}_1)(\lam-\ovl{\Lam}_2)
  \\
  \ovl{\Lam}_{1,2} &= \frac{E'}{2E}\left[
                   1 \mp \sqrt{1-\frac{2E}{\cc}} \right]
  \qquad (\ovl{\Lam}_1 < \ovl{\Lam}_2) \:.
%
\end{align}
Here 
$ E' = E -\cc\mag^2/2$.
Note that $\ovl{A}:=-8\cc E/\hbar^2\propto -c_2^2 <0$ 
for $c_2<0$ as well as for $c_2>0$, and in both cases we will have
$\ovl{\Lam}_1 \leq \lam(t) \leq \ovl{\Lam}_2$.
Following a procedure analogous to that above, we arrive at
\begin{equation}
  \lam(t) = \ovl{\Lam}_1 + 
  (\ovl{\Lam}_2-\ovl{\Lam}_1)\sin^2\!\left( 
    \ovl{\gamma}_0 - \frac{\pi}{T_{B=0}}(t-t_0) \right)
  \label{eq:rhotnoB}
\end{equation}
with
$ \sin\ovl{\gamma}_0 = 
\left[(\lin-\ovl{\Lam}_1)/(\ovl{\Lam}_2-\ovl{\Lam}_1)\right]^{1/2} $.
In this case, $\lam$ follows a pure sinusoidal evolution 
as has been predicted before in a number of references,
e.g.,~\cite{Moreno2007,Zhang2005,Pu1999}.
The average value is 
$ \lam^\textrm{av} 
 = (\ovl{\Lam}_1+\ovl{\Lam}_2)/2
 = E'/(2E) 
$, and
the period reads (compare with~\cite{Pu1999})
\begin{equation}
  \label{eq:period2}
  T_{B=0} = \frac{\pi\hbar}{\sqrt{2\cc E}} \:.
\end{equation}

\begin{figure}
  \centering
  \includegraphics[width=0.4\textwidth,clip=true]{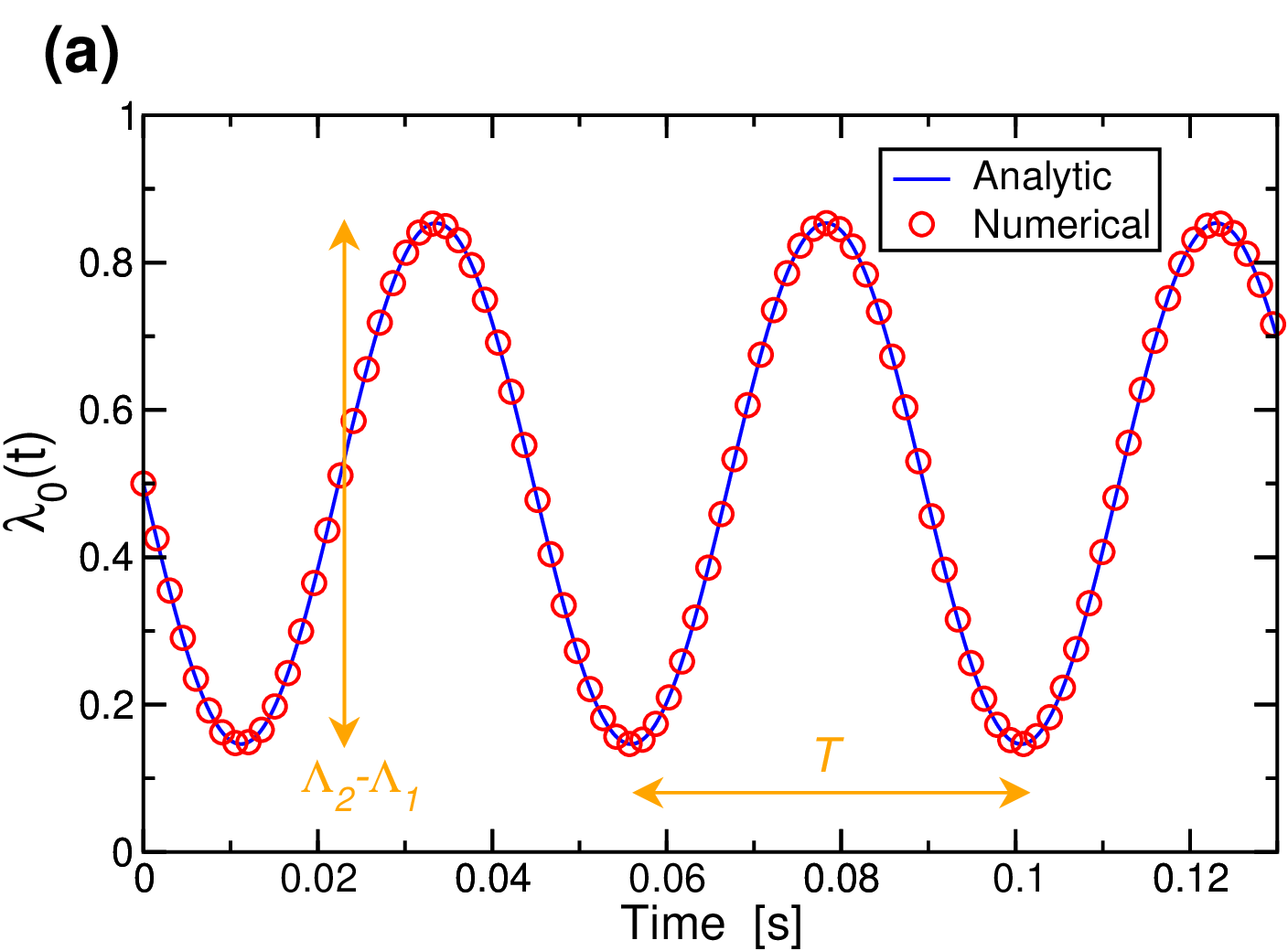}\hfill%
  \includegraphics[width=0.4\textwidth,clip=true]{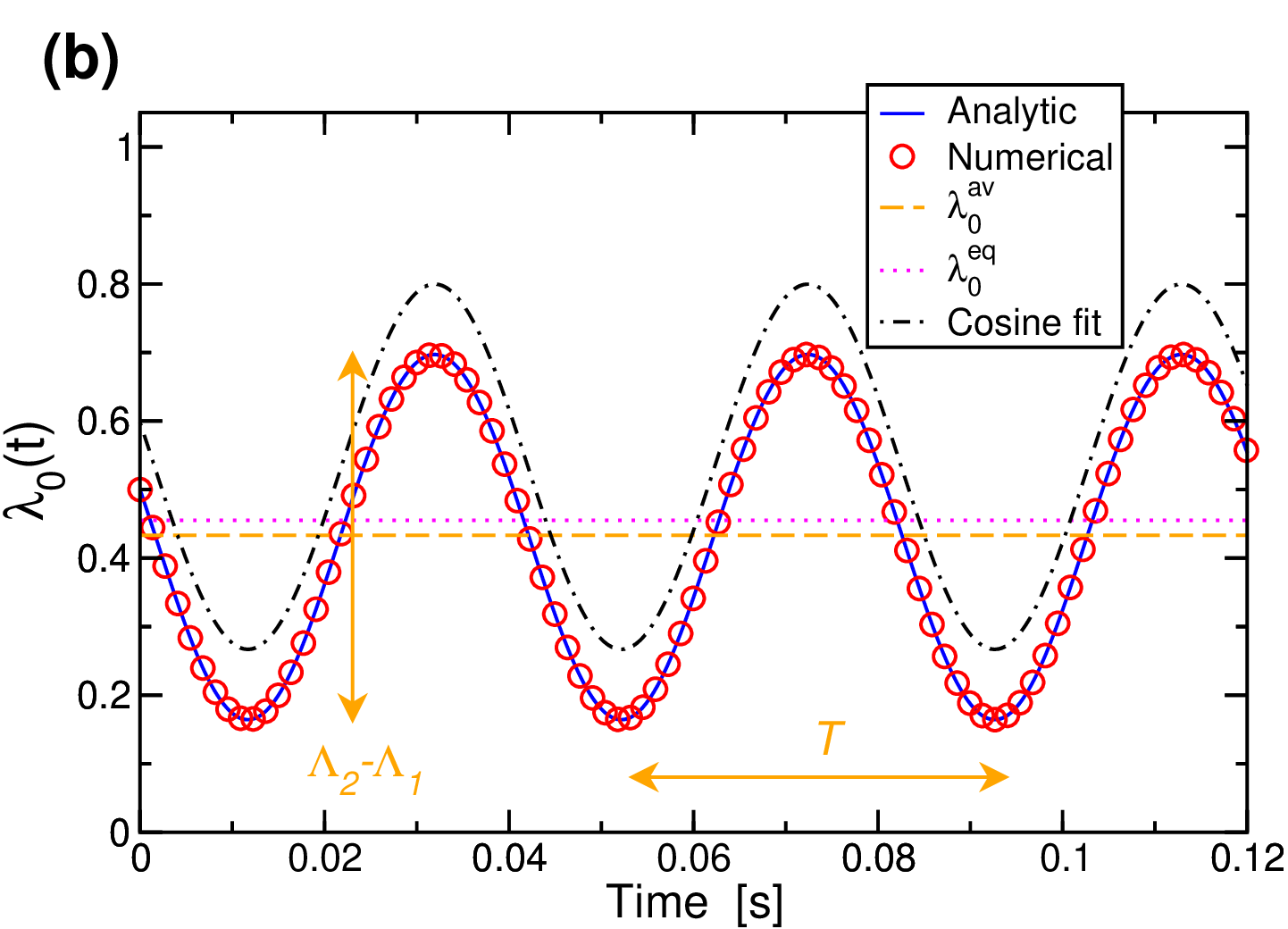}
  \caption{
    Evolution of the population of the $m=0$ Zeeman component,
    $\lam(t)$, for the cases
    (a) $\mag=0,~ B=0$,
    and (b) $\mag=0.3,~ B=100$~mG,
    starting in both instances from $\lin=0.5,~\thin=\pi/2$.
    In both panels, the solid line corresponds to the analytic
    result, (a) Eq.~(\ref{eq:rhotnoB}) or (b)~(\ref{eq:rhot}), while
    the circles are a numerical integration 
    of the differential equation for $\ldot$. The dashed line gives the
    expected average value of $\lam$, $\lam^\textrm{av}\approx 0.433$.
    The arrows indicate the amplitude and period as predicted by the
    analytical results.
    In the bottom plot, also the value of the equilibrium population
    $\leqq\approx 0.455$ is indicated by a dotted line, while the
    dashed-dotted line stands for a fit to Eq.~(\ref{eq:sinus})
    (displaced vertically by 0.1 for clarity).}
  \label{fig:rhot}
\end{figure}

\begin{figure}
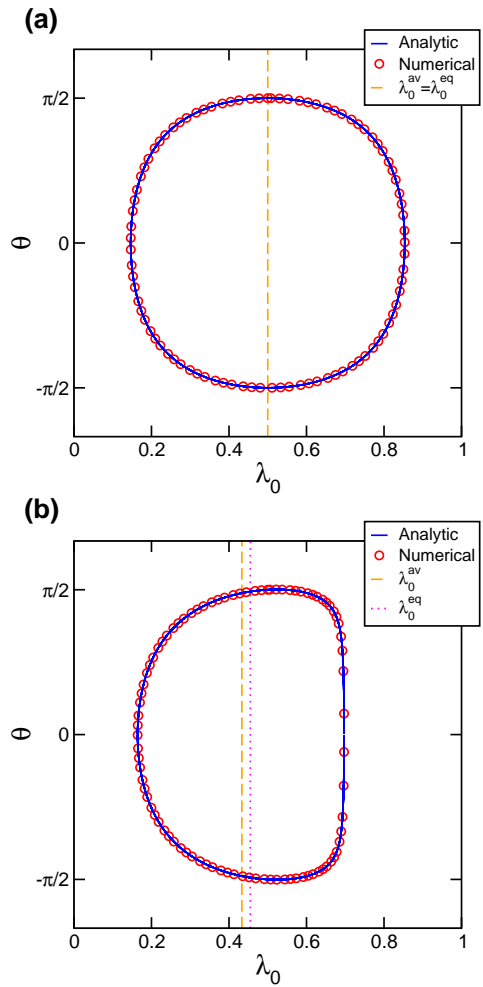

  \centering
  \includegraphics[width=0.35\textwidth,clip=true]{evol_anal_1_orbita}\hfill%
  \includegraphics[width=0.35\textwidth,clip=true]{evol_anal_2_orbita}
  \caption{
    Phase-space plot $(\lam,\th)$ corresponding to 
    {\bf (a)} the evolution shown in Fig.~\ref{fig:rhot}(a) and 
    {\bf (b)} Fig.~\ref{fig:rhot}(b) (compare with
    Fig.~\ref{fig:orbites}).
    The solid line is the analytic result in (a) Eq.~(\ref{eq:rhotnoB})
    and (b) Eq.~(\ref{eq:rhot}), 
    while the circles are the solution of the differential
    equations for $\ldot$ and $\dot{\th}$. The vertical dashed lines
    stand for the average value $\lam^\textrm{av}$ in each case.}
  \label{fig:orbites2}
\end{figure}

We show in Fig.~\ref{fig:rhot} the time evolution of $\lam(t)$ for two
representative cases. The different panels compare the analytic
evolution --given by Eq.~(\ref{eq:rhot}) or~(\ref{eq:rhotnoB})--
with a numerical solution of the corresponding equation for
$\ldot$. In all cases, we see that the amplitude as well as the period
of the time evolution are well predicted by the analytic results.
Finally, we show in Fig.~\ref{fig:orbites2} a plot of $\th(t)$
vs. $\lam(t)$ corresponding to the time evolution depicted in
Fig.~\ref{fig:rhot}. For the case with magnetic field and $\mag\neq0$
we observe that the average value $\lam^\textrm{av}\approx0.433$
(indicated by the dashed line) differs from the position of the
minimum of $\cal E$ ($\leqq\approx0.455$, cf.~Fig.~\ref{fig:orbites})
due to the deformation of the orbit.

\section{Dynamics of the trapped system}
\label{sec:dynam}

We have established in the previous section the dynamical evolution of
a homogeneous spin-1 condensate, in terms of 
orbits in the $(\lam,\th)$ plane constrained by (i) conservation of
density, (ii), conservation of magnetization, and (iii) conservation
of energy. 
The resulting dynamics of the population of the $|m=0\rangle$ Zeeman
component has been shown to be a periodic function of time, with a
period determined by the density $\dens$ of the system,
its magnetization $\mag$, as well as the initial conditions of the
evolution (implicit in $E$ and, therefore, in $\{ \Lam_j \}_{j=1,2,3}$ or
$\{\ovl{\Lam}_j\}_{j=1,2}$), cf.~Eqs.~(\ref{eq:period})
and~(\ref{eq:period2}).
Now, we will transfer these results to a realistic case of a
trapped, quasi-one-dimensional (1D) condensate.

\subsection{Local-density approximation}
\label{ssec:lda}

The initial conditions for the evolution of a trapped spinor
condensate are the set of complex values $\psi_m^\textrm{in}(\x)$ for
all Zeeman components $m$ and all positions $\x$ where the density is
not zero. 
In typical experiments, the preparation of the initial state is such
that $\dens_m(\x)/\dens(\x)$ is a constant independent of position. 
This, together with the fact that $|c_2|\ll |c_0|$ for the systems
studied so far, has lead to some theoretical works based on the
so-called single-mode approximation (SMA), which assumes 
that $\dens_m(\x,t)=\dens(\x)\lambda_m(t)$ for all times $t$ of the
evolution, i.e., that the spatial variation in the density of each
Zeeman component is always given by the total density profile.
However, numerical studies beyond the SMA
(e.g.,~\cite{Mur2006,Pu1999,Saito2005}) predicted the formation of 
{\em spin domains} as time goes by. 
These have been observed in a number of experiments,
e.g.,~\cite{Stenger1998,Higbie2005}.
In order to be able to observe the formation of spin domains during
time evolution in a trapped system, we will therefore not make use of
the SMA, but apply the analytical results of Sect.~\ref{sec:homog}
{\em via} the LDA, i.e., we will assume
that the evolution of the $|m=0\rangle$ population at each point
within the condensate, $\lam(\x,t)$, is given by Eq.~(\ref{eq:rhot})
[or Eq.~(\ref{eq:rhotnoB})]
with the substitution $\dens \rightarrow \dens(\x)$.
Here, the total density is normalized to the total number of atoms in
the condensate, $\int d\x\, \dens(\x) = N$.
Similarly, we introduce the local densities of atoms in a
given Zeeman state $\dens_m(\x,t)$ normalized as
$\int d\x\, \dens_m(\x,t) = N_m(t)$. 
The conservation laws read now $\sum_m N_m(t)=N$ and
$\sum_ m mN_m(t) = \mag$. 
We note that $\dens(\x)$ does not change in time at low enough
temperatures~\cite{Mur2006} unless momentum is imparted to the
center of mass or to one or more of the Zeeman
components~\cite{Guilleumas2008}.

In the language of the phase space introduced in
Sect.~\ref{ssec:static}, a trapped system corresponds to an
infinite-dimensional phase space, with a pair of
variables $(\lam(\x),\th(\x))$ associated to each point $\x$.
According to the LDA, we divide this whole phase space in sections
corresponding to the different positions,
and assume that they are independent.
The initial condition described above, $\dens_m(\x)/\dens(\x)=
\textrm{const}$,  corresponds then to the dynamical system starting
in {\em all} the different positions $\x$ at the same point of the
corresponding phase space, $(\lam(\x,t=0)=\lin, ~\th(\x,t=0)=\thin)$.
The dynamical evolution of the system corresponds then to the
population $\lambda_0(\x,t)$ at each point $\x$
following its own particular orbit in the 
corresponding $(\lam(\x),\th(\x))$ space, 
that is, $\lam$ at position $\x$ follows the dynamical equation 
of the homogeneous system~(\ref{eq:rhot}) [or Eq.~(\ref{eq:rhotnoB})] 
with the parameters $\Lam_j$ and $A$ determined by the local density
$\dens(\x)$.
In other words, we assume that the position dependence is only
parametric, and comes through the values of the parameters
$\Lam_j=\Lam_j(\x)$ and $T=T(\x)$. 
We will indicate this by
$\lam(\x,t) = \lam^\textrm{LDA}(\x,t) \equiv \lam^{\dens(\x)}(t)$.
%
The density at position $\x$ of atoms in the Zeeman component $m$ at
time $t$ will then be
\begin{equation}
  \dens_m(\x,t) = \dens(\x)\lambda_m(\x,t) \:.
\end{equation}
with $\lambda_{\pm1}(\x,t)=\lambda_{\pm1}^{\dens(\x)}(t)$ given by
Eq.~(\ref{eq:lambdes}) with the substitution
$\lam \rightarrow \lam^{\dens(\x)}(t)$, and $\mag=\mag(t=0)$ is a
conserved quantity~\cite{Mur2006}.

Note that the orbits associated to different points $\x$ may differ
from one another, as their shapes depend {\em inter alia} on the local
density $\dens(\x)$, cf.~Eq.~(\ref{eq:Ehomog}).
This fact, together with the position dependence of the parameters
$\Lam_j(\x)$ and $T(\x)$, is
expected to lead to a dephasing of the evolution of the partial
densities $\dens_m(\x,t)$ at the different 
points, washing out the oscillations in the integrated populations,
$N_m(t)$, in contrast to the stable oscillations that we have found
for the homogeneous system, cf.~Fig.~\ref{fig:rhot}.

In order to evaluate $N_m(t)$ it is necessary to know the density
profile of the system. A good estimate for $\dens(\x)$ in trapped
atomic gases is given by the Thomas-Fermi approximation,
\begin{align}
  \dens_\textrm{TF}(\x) &=
  \left\{ \begin{array}{ll}
      \frac{\dens_\textrm{max}}{R_\textrm{TF}^2}(\Rtf^2-|\x|^2) \:, 
      & \quad |\x| \leq \Rtf \\
      0 \:, & \quad \textrm{otherwise}
    \end{array} \right. \:.
  \label{eq:tf}
\end{align}
For a quasi-1D system with total number of atoms $N$ and central
density $\nmax$, $\Rtf=3N/(4\nmax)$. The integrated population in
$|m=0\rangle$ then reads
\begin{equation}
  N_0(t) = \int dx\,
  \dens_\textrm{TF}(x)\lam^{\dens_\textrm{TF}(x)}(t) \:.
  \label{eq:integral}
\end{equation}

\subsection{Analytic approximation with sinusoidal time dependence}
\label{ssec:sinus}

The time dependence of $\lam^{\dens(x)}(t)$ has in principle to be
calculated from Eq.~(\ref{eq:rhot}) for each position $x$ at each time
step, and then the integral~(\ref{eq:integral}) performed
numerically to determine $N_0(t)$.
It is possible however to give an analytical estimation for $N_0(t)$
if we make a further assumption on the time evolution.
From Fig.~\ref{fig:rhot}, we see that the evolution of $\lam(t)$ for
the homogeneous system is very close to a sinusoidal function even
when $B\neq 0$~%
\footnote{Only for fields close to the resonance,
  $c_2\dens\simeq\delta$, does the time evolution of the homogeneous
  system differ notably from a sine function, cf.~\cite{inprogress}.%
}%
. This is illustrated in Fig.~\ref{fig:rhot}(b), where a
function of the form 
\begin{equation}
  \label{eq:sinus}
  \lam^\textrm{cos}(t) = a + b\cos(\gamma+\nu t)
\end{equation}
has been fitted to the numerical values obtained from
Eq.~(\ref{eq:rhot}). The fit is very good, even for this case, where
the orbit in phase space is strongly deformed
[cf.~Fig.~\ref{fig:orbites2}(b)]. The advantage of approximating the
time evolution of $\lam$ by Eq.~(\ref{eq:sinus}) is that it allows for
an analytic evaluation of the spatial integral~(\ref{eq:integral}),
taking into account the position dependence of $\nu$. Indeed, from
Eq.~(\ref{eq:period}) we expect
$\nu(x) \propto \dens(x) \propto (\Rtf^2-x^2)$.
It is easy to show that
\begin{align}
  N_0^\textrm{cos}(t)
  &= \int dx\, \dens_\textrm{TF}(x) 
  \left[a + 
    b\cos\!\left[\gamma 
      +\nu_0(1-\frac{x^2}{\Rtf^2})t\right] \right]
  \nonumber \\
  &=\frac{\dens_\textrm{max}}{6(\tt)^{3/2}}\Big[
    8a(\tt)^{3/2}
    +6b\sqrt{\tt}\sin(2\gamma) \nonumber \\
  & +3\sqrt{2\pi}b \left\{ 
       \cos(\mu)S(\eta) - \sin(\mu)C(\eta) \right\}
  \nonumber \\
  &  +6\sqrt{2\pi}b\tt \left\{
       \cos(\mu)C(\eta) + \sin(\mu)S(\eta) \right\}
    \Big] .
  \label{eq:int_sinus}
\end{align}
Here 
$S(\eta)$ and $C(\eta)$
are the Fresnel integrals~\cite{Abramowitz}, and we
introduced $\mu=\gamma+\tt$ and $\eta=\sqrt{2\tt/\pi}$.

\begin{figure}
  \centering
  \includegraphics[width=0.45\textwidth,clip=true]{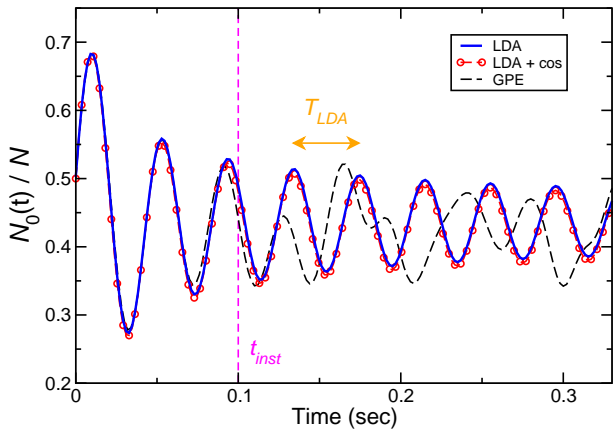}
  \caption{
    Time evolution of the integrated $|m=0\rangle$ population
    (normalized to the total population, $N=20\,000$) for 
    $\lin=0.5, ~\thin=\pi/2, ~\mag=0.3$ and $B=100$~mG.
    The solid line shows the LDA result, Eq.~(\ref{eq:integral}), with
    $\lam(x,t)=\lam^{\dens(x)}(t)$. The circles stand for the analytic
    estimate of Eq.~(\ref{eq:int_sinus}), and the dashed line is the
    result of integrating the set of coupled Gross-Pitaevskii
    equations~(\ref{eq:gpe}).}
  \label{fig:compar}
\end{figure}

We show in Fig.~\ref{fig:compar} the time evolution of the integrated
$|m=0\rangle$ population as given by Eqs.~(\ref{eq:integral})
and~(\ref{eq:int_sinus}). 
This calculation has been done for a quasi-1D system of $20\,000$
$^{87}$Rb atoms in a trap such that the central density is
$4.4\times10^{14}$~cm$^{-3}$.
The initial conditions are $\lin=0.5, ~\thin=\pi/2$ and $\mag=0.3$ and
we have taken a magnetic field $B=100$~mG
(cf.~Fig.~\ref{fig:orbites}).
The solid line in the figure corresponds to the numerical integration
of Eq.~(\ref{eq:integral}) with $\lam(x,t)=\lam^{\dens(x)}(t)$ given
by Eq.~(\ref{eq:rhot}). The circles stand for the analytic
expression~(\ref{eq:int_sinus}) with the parameters
$a,b,\gamma,\nu_0$ taken so that $\lam^\textrm{cos}(t)$ for
a homogeneous system with density $\dens=\nmax$ reproduces the
same behavior as that given by Eq.~(\ref{eq:rhot}) at the same
density: 
$a=(\Lam_1+\Lam_2)/2, ~b=(\Lam_1-\Lam_2)/2, ~\gamma=2\gamma_0,
~\nu_0=2\pi/T_\textrm{LDA}$, and $T_\textrm{LDA}=T(\nmax)$. 
The agreement between the two calculations is very good at all times.
Therefore, we conclude that the average value of $\lam$ as well as the
characteristic period of the oscillations is well determined by the
values $\Lam_j$ and $T_\textrm{LDA}$ 
calculated with the central density, while the time scale for the
damping of the oscillations is determined by the spatial profile of
the density.

Regarding the dephasing of the evolution of $\lam(\x,t)$ among
different points, it is not very strong, in the sense that the damping
of the oscillations is relatively slow. To be more precise, one can
have a reasonable fit 
to the solid line in Fig.~\ref{fig:compar} by a function of the form
\begin{equation}
  \label{eq:nonexp}
  N_0(t) = \ovl{N}_0 +
  \Delta N_0 \exp(-\al\sqrt{t}) 
  \cos\left(2\gamma_0'+\frac{2\pi}{T'}t\right) 
\end{equation}
with $\ovl{N}_0\approx a$, 
$\Delta N_0 \approx b$, 
$\gamma_0' \approx \gamma_0$
and $T' \approx T_\textrm{LDA}$.

\subsection{Comparison with the mean-field approach}
\label{ssec:gpe}

We proceed finally to compare the approximate calculation of $N_0(t)$
with a more complete approach in terms of the dynamical equations for
the three components of the vector order parameter, 
$\psi_m(\x,t)$, cf.~Eq.~(\ref{eq:order}). In the mean-field
approximation, such equations can be cast in the form of three coupled
Gross-Pitaevskii equations~\cite{Mur2006},
\begin{subequations}
  \begin{align}
%
    i \hbar \frac{\partial \psi_{\pm 1}}{\partial t}
    =&
    [{\cal H}_s  + c_2(\dens_{\pm 1}+\dens_0- \dens_{\mp 1})] \psi_{\pm 1}
    + c_2 \psi_0^2 \psi^{*}_{\mp 1} \,, \\
    i \hbar \frac{\partial \psi_0}{\partial t} =&
    [{\cal H}_s  + c_2(\dens_{1}+\dens_{-1})] \psi_{0} 
    + 2c_2 \psi_{1} \psi_0^* \psi_{-1} \,,
  \end{align}
  \label{eq:gpe}
\end{subequations}
where
${\cal H}_s=-\hbar^2/(2M)\, {\bm \nabla}^2 +V_{\textrm{ext}}(\x)+c_0\dens(\x)$
and $c_0=4\pi\hbar^2(a_0+2a_2)/(3M)$.

The results of solving Eqs.~(\ref{eq:gpe}) with a Runge-Kutta
algorithm are included in Fig.~\ref{fig:compar} as a dashed line.
The average value of the oscillating $\lam(t)$ is well
estimated by the analytical model of Sect.~\ref{ssec:dynam}.
Also, the characteristic time scale of the oscillations is well
estimated by Eq.~(\ref{eq:period}).
The overall agreement is good for times $t\lesssim 100$~ms.
After this time, the analytical estimate keeps oscillating with a
slowly decreasing amplitude, while the numerical solution of the
coupled equations~(\ref{eq:gpe}) shows fluctuating oscillations.
This behavior has been observed before, and the transition at
$t=t_\textrm{inst}\sim100$~ms has been related to a dynamical
instability that leads to the formation of dynamical
spin domains in the system~\cite{Saito2005,Mur2006,Higbie2005}.
It is thus not surprising that our simple model fails for
$t\gtrsim t_\textrm{inst}$.
It is nevertheless remarkable that the time scale set by
$T_\textrm{LDA}=T(\nmax)\approx89$~ms is still a good estimate of the
characteristic oscillation time even much later during the time
evolution.

\section{Dephasing in a magnetic field and the process of 
  structure formation in finite systems}
\label{sec:dephasing}

A qualitative difference between the homogeneous system and the
confined one appears when a magnetic field is present and, therefore,
$A\neq0$. The dynamics of a spinor condensate in a magnetic field is
known to show two limiting behaviors: the mean-field regime, where
the interaction energy dominates the evolution, and the Zeeman
regime, where the evolution is driven by the Zeeman term of the
Hamiltonian~\cite{Zhang2005,Sadler2006,Kronj2006}. The crossover
between the two regimes occurs when $c_2n \sim \delta$. This
transition can be studied in real time by changing the (homogeneous)
magnetic field on which the condensate is 
immersed~\cite{Sadler2006,Saito2007,inprogress}.

This transition can also be
observed between different spatial regions of an inhomogeneous
system. Indeed, if we assume that the magnetic field, magnetization
and central density are chosen so that $|c_2|\dens(x=0) > \delta$
(so that at the center we are in the mean-field regime), then at the
wings of the system, where $\dens(x)\rightarrow 0$, we will be in the 
Zeeman regime.
%
Therefore, we expect to have a region in real space where the
behavior with time changes qualitatively. For a profile as in 
Eq.~(\ref{eq:tf}), this transition border is given by 
\begin{equation}
  \frac{x_\textrm{trans}}{\Rtf}
  = \textrm{Re} 
  \left[ 1 - \frac{\delta}{c_2\nmax} \right]^{1/2} \:.
  \label{eq:trans}
\end{equation}
Naturally, for $\delta=0$, there is no transition (the density
vanishes at $x=\Rtf$).
On the other hand, for large enough magnetic field 
the whole system is in the Zeeman regime ($x_\textrm{trans}=0$).

These two regimes evolve with different characteristic times,
$\Tmf\simeq\hbar/(|c_2|\dens)$ and
$\TZ\simeq\hbar/\sqrt{2c_2\dens\delta}$,
cf.~Eqs.~(\ref{eq:period2}) and~(\ref{eq:period}).
Because of this, we can expect $\lam$ and the phase in the inner part of
the condensate ($|x|<x_\textrm{trans}$) to evolve at a different rate
than in the outer wings of the system ($|x|>x_\text{trans}$),
resulting in a particular spatial dependence of the phase.
We note that the appearance of a spatial structure in the phase will
lead to the creation of spin currents~\cite{Mur2006} and, thus,
to spin textures as reported in~\cite{Chang2005,Sadler2006}.
Even though a smooth density profile will lead to a smooth variation
in $T(x)=T(n(x))$ with position, from our model we expect that these
qualitatively different behaviors should be observable for times
$t \gtrsim t_\textrm{trans} =
 \textrm{min}\{\Tmf, \TZ\}$

Interestingly, in light of the discussion in Sect.~\ref{ssec:gpe},
we observe that the time when the dynamical instability is expected to
set in is close to the time when the divergence between mean-field and
Zeeman regimes should be observable:
$t_\textrm{inst}\simeq t_\textrm{trans}$.
Because processes such as spin currents fall beyond LDA, their
appearance implies a breakdown of our model, which is therefore not
applicable to analyze the process of structure formation. 
This breakdown explains the lack of agreement between the results of
our LDA model and those from Eqs.~(\ref{eq:gpe}) for
$t \gtrsim t_\textrm{inst}$ observed in Fig.~\ref{fig:compar}.


The experiments reported in Ref.~\cite{Sadler2006} showed the
appearance of spin domains to be simultaneous with that of topological
defects (phase windings) and also spin currents. This observation is
consistent with the model just sketched. The time scale for the
appearance of spin domains is estimated in that reference to be
$\sim \hbar/(2|c_2|\dens)$~%
\footnote{We observe an apparent misprint on page 313 of
  Ref.~\cite{Sadler2006}, where the time scale is written as
  $\hbar/\sqrt{2|c_2|\dens}$,which is not dimensionally consistent.}%
.
Similarly, Saito {\em et  al.}~\cite{Saito2007} determined the
time scale for the occurrence of a dynamical instability to be
$t_\textrm{inst}=\hbar/(|c_2|\dens)$ when the magnetic field is small;
this estimate coincides with our $\Tmf$. 
On the other hand, for larger magnetic fields
[$q\geq |c_2|\dens$ with $q=(\mu_B B)^2/(4E_\textrm{hfs})$],
the relevant instability time scale is
$t_\textrm{inst}=\hbar/\sqrt{q|q+2c_2\dens|}$, which is similar to 
$\TZ$.

From their simulations, Saito and Ueda indicated~\cite{Saito2005}
that the formation of spin domains starts at the center of the
condensate, and then spreads out. In our model, however, the position
where the phase slip appears is determined by $x_\textrm{trans}$, and
therefore is in principle amenable to be modified experimentally.
It seems interesting to investigate the prospect to control
the spatial appearance of spin domains and phase structures as
predicted by Eq.~(\ref{eq:trans}).


\section{Summary and conclusions}
\label{sec:conclusions}

We have studied the dynamics of a trapped spin-1 condensate under a
magnetic field. 
First, we have analyzed the homogeneous system and seen that its
dynamics can be understood in terms of orbits in the $(\lam,\th)$
space. We have then solved analytically for the dynamical evolution
$\lam(t)$. We have used this information to study the trapped system by
means of the Local Density Approximation (LDA). The results of this
approach agree with those of the mean-field treatment for evolution
times before the occurrence of a dynamical
instability~\cite{Saito2005}. In particular, the expected average
value of $\lam$, as well as the characteristic time scale of its
dynamics are well predicted by the formulas for the homogeneous
system.

Our analysis of the trapped system has shown that, in the
presence of a magnetic field, both the mean-field and Zeeman
regimes are realized in a single spinor condensate.
The analysis of this model allows for some qualitative insight into
the process of structure formation.
In particular, our model identifies a transition point
[cf. Eq.~(\ref{eq:trans})] around which this structure is generated,
and predicts that it should be tunable, which could be tested in
future experiments.

\begin{acknowledgments}
I would like to acknowledge A. Polls, A. Sanpera, and T. Petit 
for discussions and encouragement to develop this project, and 
M. D. Lee for comments on an early version of the paper.
This work was supported by the UK EPSRC (grant no.~EP/E025935).
\end{acknowledgments}

\bibliographystyle{apsrev}
\bibliography{bibliospin}

\end{document}